\documentclass[a4paper]{article}

\usepackage[english]{babel}
\usepackage[utf8x]{inputenc}
\usepackage[T1]{fontenc}
\usepackage{setspace}
\usepackage{graphicx}
\usepackage{caption}
\usepackage{subcaption}
\usepackage[labelfont=bf]{caption}
\usepackage{multicol}
\usepackage[a4paper,top=3cm,bottom=3cm,left=3cm,right=3cm,marginparwidth=1.75cm]{geometry}

\usepackage{gensymb}
\usepackage{amsmath}
\usepackage{graphicx}
\usepackage[colorinlistoftodos]{todonotes}
\usepackage[colorlinks=true, allcolors=blue]{hyperref}

\title{Optimizing Thermochromism of Solution-Processed VO$_2$ Nanocomposite Films for Chromogenic Fenestration}
\date{July 29, 2017}
\author{Joel P. Abraham\footnote{Westwood High School, Austin, TX, USA, 78750} \footnote{This research was conducted at The University of Texas at Austin under the mentorship of Lauren Gilbert and Dr. Delia J. Milliron}}

\begin{document}
\maketitle

\begin{abstract}
Vanadium (IV) oxide is one of the most promising materials for thermochromic films due to its unique, reversible crystal phase transition from monoclinic (M1) to rutile (R) at its critical temperature (T$_c$) which corresponds to a change in optical properties: above T$_c$, VO$_2$ films exhibit a decreased transmittance for wavelengths of light in the near-infrared region. However, a high transmittance modulation often sacrifices luminous transmittance which is necessary for commercial and residential applications of this technology. In this study, we explore the potential for synthesis of VO$_2$ films in a matrix of metal oxide nanocrystals, using In$_2$O$_3$, TiO$_2$, and ZnO as diluents. We seek to optimize the annealing conditions to yield desirable optical properties. Although the films diluted with TiO$_2$ and ZnO failed to show transmittance modulation, those diluted with In$_2$O$_3$ exhibited strong thermochromism. Our investigation introduces a novel window film consisting of a 0.93 metal ionic molar ratio VO$_2$-In$_2$O$_3$ nanocrystalline matrix, demonstrating a significant increase in luminous transmittance without any measurable impact on thermochromic character. Furthermore, solution-processing mitigates costs, allowing this film to be synthesized 4x-7x cheaper than industry standards. This study represents a crucial development in film chemistry and paves the way for further application of VO$_2$ nanocomposite films in chromogenic fenestration.

\end{abstract}

\doublespacing
\section{Introduction}

An estimated \$35 billion of energy is lost annually through windows with 45\% of the average American household energy bill spent on heating and cooling. Window films offer a cost-effective method to reduce this wasted energy, functioning as insulators by minimizing heat transfer through the window. Conventionally, windows are double glazed and a low-emissivity coating is applied to ensure a low solar heat gain, but recent developments in film chemistry have enabled chromogenic fenestration — thermochromic films with reflective properties that vary with an applied stimulus, thus minimizing energy loss.

\begin{figure*}[t]
    \centering
    \includegraphics[scale=.45]{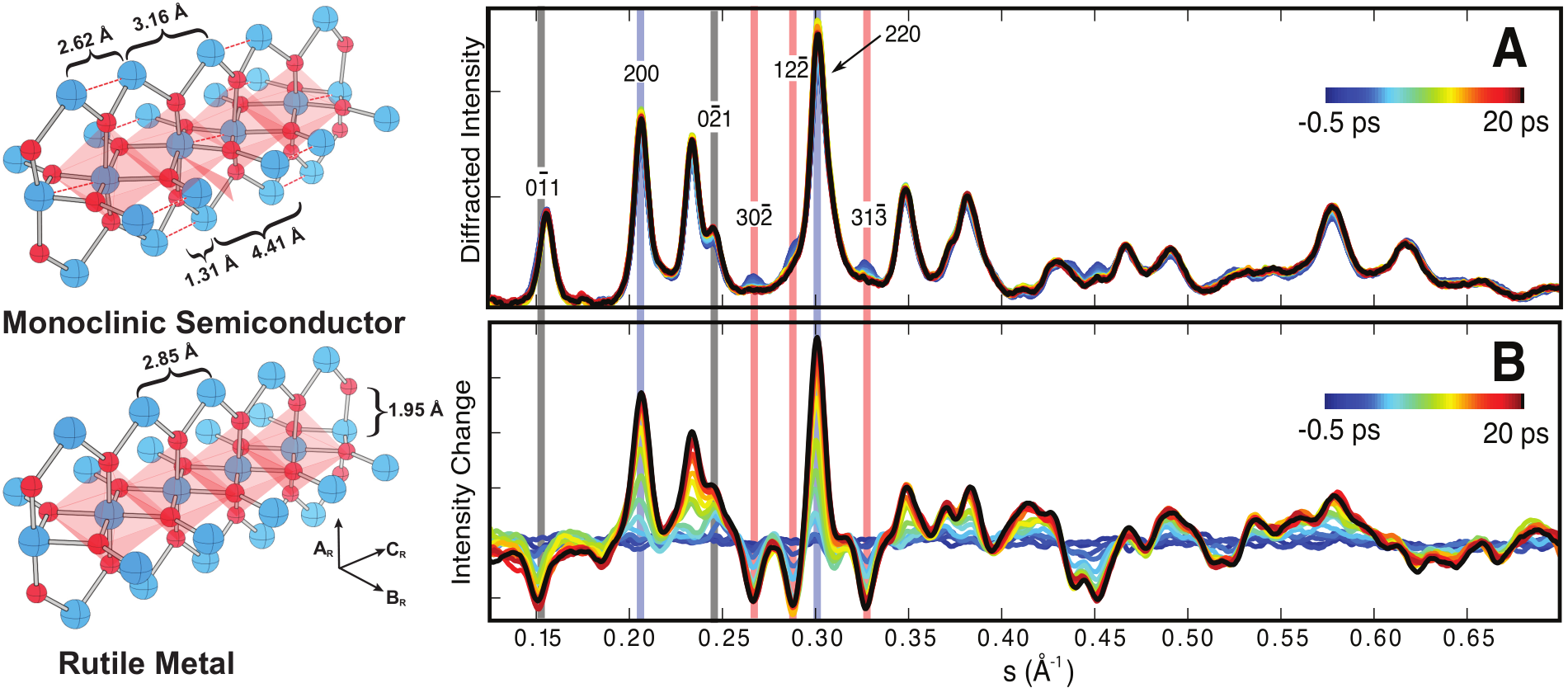}
    \caption{\textbf{VO$_2$ Crystal Morphology} \cite{siwick}}
    \label{vo2}
\end{figure*}

Vanadium (IV) oxide is one of the most promising materials for thermochromic films due to its unique, reversible crystal phase transition from monoclinic (M) to rutile (R) at its critical temperature (T$_c$) — approximately $68\degree $C for undoped bulk VO$_2$ (Figure \ref{vo2}). \cite{morin} Corresponding to the structural transformation, VO$_2$ undergoes a change in its electronic character from semiconducting to metallic which translates to a change in optical properties. This metal-insulator transition (MIT) has long been known to be associated with the dimerization of Vanadium ions caused by Peierls instability, but early researchers struggled to describe exactly what role the electronic transition played in this classic Mott transition caused by strong Coulomb repulsion. \cite{qazilbash} Certain evidence suggests that the MIT is a product of the interaction between the structural and electronic transitions while other evidence suggests that the structural and electronic transitions are not linked but overlaid, and the electronic transition is the driving mechanism behind the VO$_2$ MIT. \cite{driscoll}

The MIT produces unique reflective properties: below T$_c$, VO$_2$ films exhibit a high transmittance for wavelengths of light in the near-infrared-region (NIR), while above T$_c$, films exhibit a low NIR transmittance. Since thermal energy from the sun is largely radiated via NIR light, this transmittance modulation can be easily exploited to minimize energy loss across windows. Pure VO$_2$ thin films have been researched extensively and their properties have been experimentally verified, but three significant obstacles remain: large cost, impractically high transition temperature, and dark color and insufficient luminous transmittance (transmittance of visible light region). \cite{li}

Thin film synthesis remains expensive and unsuitable for commercial or residential applications. Recent approaches to decrease the cost of film-deposition involve solution-processed VO$_2$ nanocrystals which are unique in their potential for low-cost, high-throughput, highly reproducible synthesis via scalable deposition techniques such as drop casting or spin coating. \cite{paik} These are preferable to epitaxial thin films which require vacuum processing, which can be very expensive and difficult to deposit. Although much of the literature has historically analyzed thin films, research in film chemistry is shifting towards nanocrystalline films since they can be synthesized more cheaply and accurately than thin films; as such, solution-processed VO$_2$ nanocrystalline films were used in this investigation.

With a transition temperature of $68\degree $C, unaltered VO$_2$ films have virtually no applications in window coatings since such a high temperature is not suitable for indoor environments. However, doping with transition metal ions with certain valencies has been demonstrated to decrease T$_c$ proportional to the concentration of the dopant. \cite{goodenough} The most frequently used dopant is W$^{6+}$ which has the potential to decrease T$_c$ to room temperature without any measurable influence on optical properties. \cite{kong}

Finally, the most crucial issue for applying this material to window coatings is the low transparency of VO$_2$ films. There exists a tradeoff between luminous transmittance and thermochromism as highly thermochromic films need to be so thick that the luminous transmittance is too low for most window applications. Current research is focused in this area; developments include layered VO$_2$ and TiO$_2$ or SiO$_2$ films and antireflective coatings but neither of these approaches have yielded significant success. \cite{jin,kakiuchida} Such a critical gap between research and application prompted an exploration into novel methods for increasing transparency of VO$_2$ films, beginning with a survey of available literature. A recent study was found, demonstrating via a computational model that nanoparticles dispersed in a dielectric host provide advantages over continuous thin films in terms of luminous transmittance and solar energy transmittance modulation. \cite{li} This source’s reliability is bolstered by the mathematical justifications for its conclusion.

This investigation is premised on the notion that diluting VO$_2$ with a white or clear metal oxide will yield an overall more transparent film with a higher luminous transmittance while preserving thermochromic character. It was hypothesized that dilution of VO$_2$ would increase luminous transmittance while not significantly mitigating transmittance modulation; although a slight decrease in transmittance modulation is expected, since the dilution of VO$_2$ implies that its optical properties will be less apparent, sufficient thermochromism is predicted to remain. The research explores the following question: how can solution-processed VO$_2$ nanocomposite films be synthesized with optimal reflective properties?

\section{Materials and Methods}
Since VO$_2$ is not readily available, it must be synthesized from V$_2$O$_3$. Once the V$_2$O$_3$ is mixed with the selected host matrix and dispersed on a film, the synthesis process involves heating the film at high temperatures (between $375\degree $C and $475\degree $C) in a chamber filled with O$_2$, as well as inert N$_2$ to control pressure, and left for an hour to properly oxidize the V$_2$O$_3$ into VO$_2$.

To address the research question, a variety of experiments were performed by varying the metal oxide host matrix, molar ratios of vanadium to host matrix metal ion, annealing temperature, oxygen partial pressure during annealing, and film thicknesses. Colloidal V$_2$O$_2$ nanocrystals were mixed with either In$_2$O$_3$, TiO$_2$, or ZnO nanocrystal solutions; these diluents were chosen since they are white or clear and abundant in the facility in which this research was conducted. Vanadium-indium samples were synthesized with $0.93$ mol V: mol In and $0.21$ mol V: mol In, while vanadium-titanium and vanadium-zinc samples had molar ionic ratios as V:Ti = $0.93$ and V:Zn = $0.93$ \footnote{To clarify this notation, A:B refers to the ratio of the moles of ions of metal A to that of metal B.}. 

Glass and silicon substrates were cut using a diamond-tipped stylus and cleaned via wash cycles of chloroform, acetone, and isopropanol; the substrates were stored in isopropanol and dried before use. Films on glass substrates were spin coated for a more uniform and thinner film, while films on silicon substrates were drop cast, ensuring a thicker coating. The drop cast process involved pipetting $60-100$ µL of solution and letting dry in the fume hood. The alternative process uses a spin coater; films of each nanocrystal mixture were spun on glass substrates under varying spin conditions, yielding varying thicknesses, although most were spun at $700$ rpm for $60$ seconds and $4000$ rpm for another $30$ seconds; these conditions were observed to yield an optimal film thickness and was kept constant for the following experimentation. Once spun, the films were stored in a vacuum desiccator to prevent oxidation. 
    
The films were then annealed in a tube furnace under varying conditions. Each film was carefully placed in a quartz tube which was secured horizontally and attached to a gas flow line on one side which flowed gas to a bubbler on the other side. O$_2$ was available at $1000$ ppm in N$_2$ from a cylinder, and was diluted with building-provided N$_2$ to yield a controlled atmosphere with a total flow of $300$ standard cubic centimeters (sccm) of gas. Oxygen partial pressures ranged from $16.667-1000$ ppm. Glass wool was used as an insulator in the tube as it was heated to high temperatures. The tube furnace was closed and film was heated at $50\degree$C for $30$ minutes then heated at a certain annealing temperature, varying between $375\degree $C and $475\degree $C, for $1$ hour. After the annealing process was completed, the lid of the Tube Furnace was opened for glass films to cool more quickly, although some Silicon films were cooled with the lid closed since it was found that rapid cooling caused the Silicon films to shatter from thermal shock. Once cooled to room temperature, the films were removed from the tubes and returned to the vacuum dessicator and the glass tubes were sealed with paraffin films to prevent contamination. 

Optical spectrometry was performed using a Linkam Cell heating stage with a UV-visible spectrometer to determine the influence of the variables on the optical properties of the films. Once the sample was placed in the cell and secured, spectra indicating the percent transmittance at each wavelength between $350$ and $2500$ nm were collected at two different temperatures, room temperature (RT) and $100\degree $C. Further analysis was performed on the samples that demonstrated modulation. For these samples, transmittance spectra were collected over a temperature series, for both the heating and cooling phases.

Other processes that were performed during this investigation, although infrequently, include X-ray Diffraction (XRD). XRD is used to identify a crystalline material and involves the diffraction of X-rays upon striking a certain crystal across varying incident angles which yields a spectra indicating relative intensity. Due to interference, certain peaks are visible at particular angles of incidence that meet the Bragg condition, and the relative peak intensities can be compared to standard reference spectra to identify the crystal.

\begin{figure*}
\centering
\begin{subfigure}{.5\textwidth}
  \centering
    \includegraphics[scale=.5]{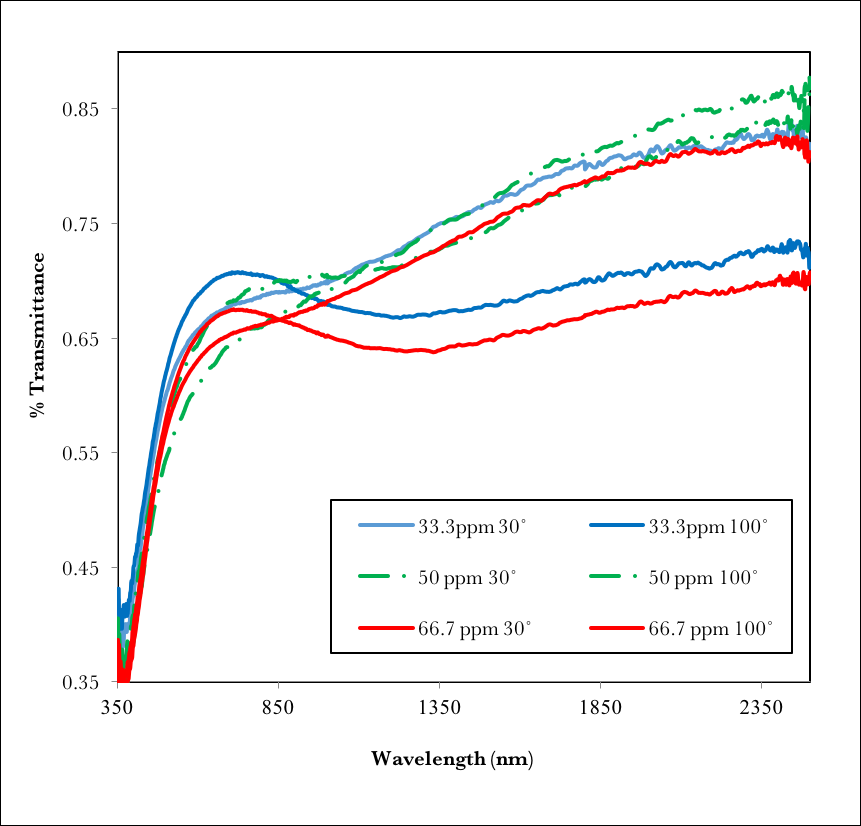}
    \caption{\textbf{375$\degree$C Spectra}}
    \label{In375}
\end{subfigure}%
\begin{subfigure}{.5\textwidth}
  \centering
    \includegraphics[scale=.5]{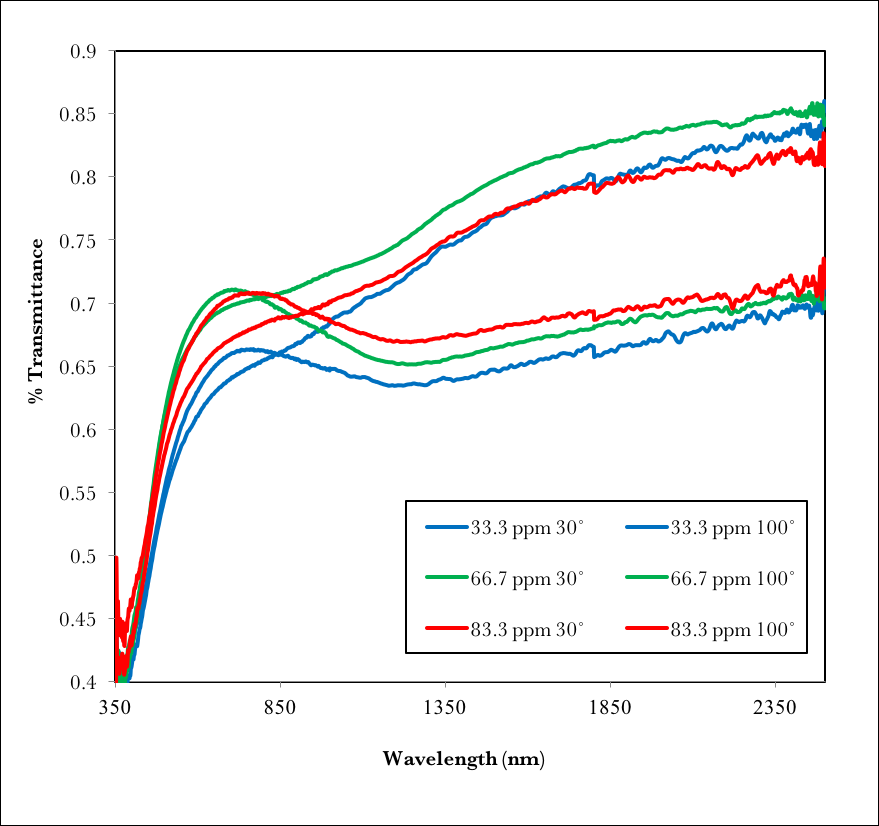}
    \caption{\textbf{400$\degree$C Spectra}}
    \label{In400}
\end{subfigure}
\caption{\textbf{0.93 VO$_x$-In$_2$O$_3$}}
\label{In}
\end{figure*}


\section{Results and Discussion}
The vanadium (III) oxide was diluted with three different diluting agents and the films were annealed, then tested for optical modulation upon heating. Only white metal oxides with high dielectric constants were used as diluting agents, since colored metal oxides would tint the film and sacrifice luminous transmittance, and high dielectric materials were computationally determined to yield greater transmittance modulation. \cite{li} Indium (III) oxide, zinc (II) oxide, and titanium (IV) dioxide were selected as the diluents, with dielectric constants of $3.35$, $8.5$, and $85$ respectively. 

The first sample tested was that of vanadium oxide diluted with indium (III) oxide, henceforth referred to as VO$_{x}$-In$_2$O$_2$\footnote{Due to the differences in the oxidation state of the vanadium oxide before and after annealing, VO$_x$ is widely used in the literature to represent an arbitrary oxide of vanadium.}. The initial annealing conditions were $375\degree$C and $83.3$ ppm O$_2$, which is similar to the conditions found to be optimal for synthesis of pure VO$_2$ films. Two batches of VO$_x$-In$_2$O$_3$ were synthesized, one with V:In = $0.93$ and the other with V:In = $0.21$. Both sets of samples were annealed at $16.7$, $33.3$, $50$, $66.7$, $83.3$, and $153.3$ ppm O$_2$. The $0.93$ VO$_x$-In$_2$O$_3$ displayed almost immediate success, as modulation was observed in the samples annealed at $375\degree $C with $33.3$ and $66.7$ ppm and $400\degree$C with $33.3$, $66.7$, and $83.3$ ppm  (Figure \ref{In}\subref{In375}, \ref{In}\subref{In400}).

There was expected to be an optimal flow condition since the samples annealed under the lowest and highest oxygen partial pressures failed to show modulation but the sample annealed at $50$ ppm also does not appear to contain VO$_2$. Although there seems to be some switching, it is not significant enough to be considered and the fact that the spectra at both $30\degree$C and $100\degree$C seem to be similar to the $33.3$ ppm and $66.7$ ppm $30\degree$C spectra suggests that a different oxide of vanadium was made. While no clear trends can be extrapolated from these data, the observation can be made that the sample annealed at $66.7$ ppm displays slightly greater modulation than the sample annealed at $33.3$ ppm but overall less visible light transmittance. 


The samples annealed at $400\degree$C all showed higher NIR modulation than those annealed at $375\degree$C in terms of transmittance modulation. The $66.7$ ppm sample exhibited slightly greater luminous transmittance than the $83.3$ ppm sample. The $66.7$ ppm sample still displayed the greatest transmittance modulation, followed by the other two samples.

\begin{figure*}[t]
\centering
\includegraphics[scale=.85]{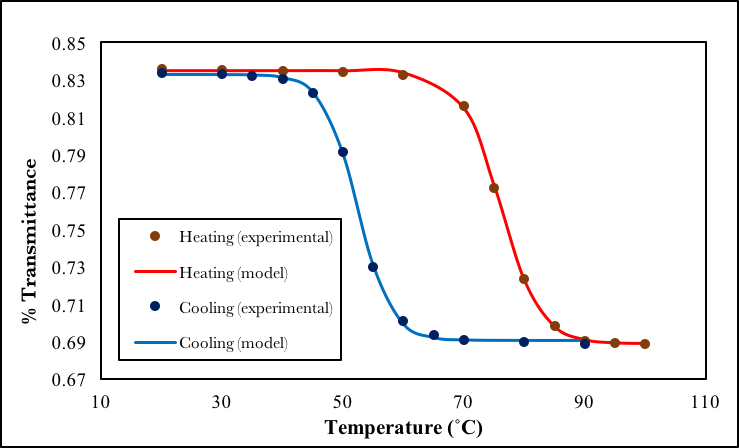}
\caption{\textbf{0.93 VO$_x$-In$_2$O$_3$ 66.7 ppm 400$\degree$ C Transmittance Hysteresis at 2000 nm}}
\label{hysteresis}
\end{figure*}

\begin{figure*}
\centering
\begin{subfigure}{.5\textwidth}
    \centering
    \includegraphics[scale=.18]{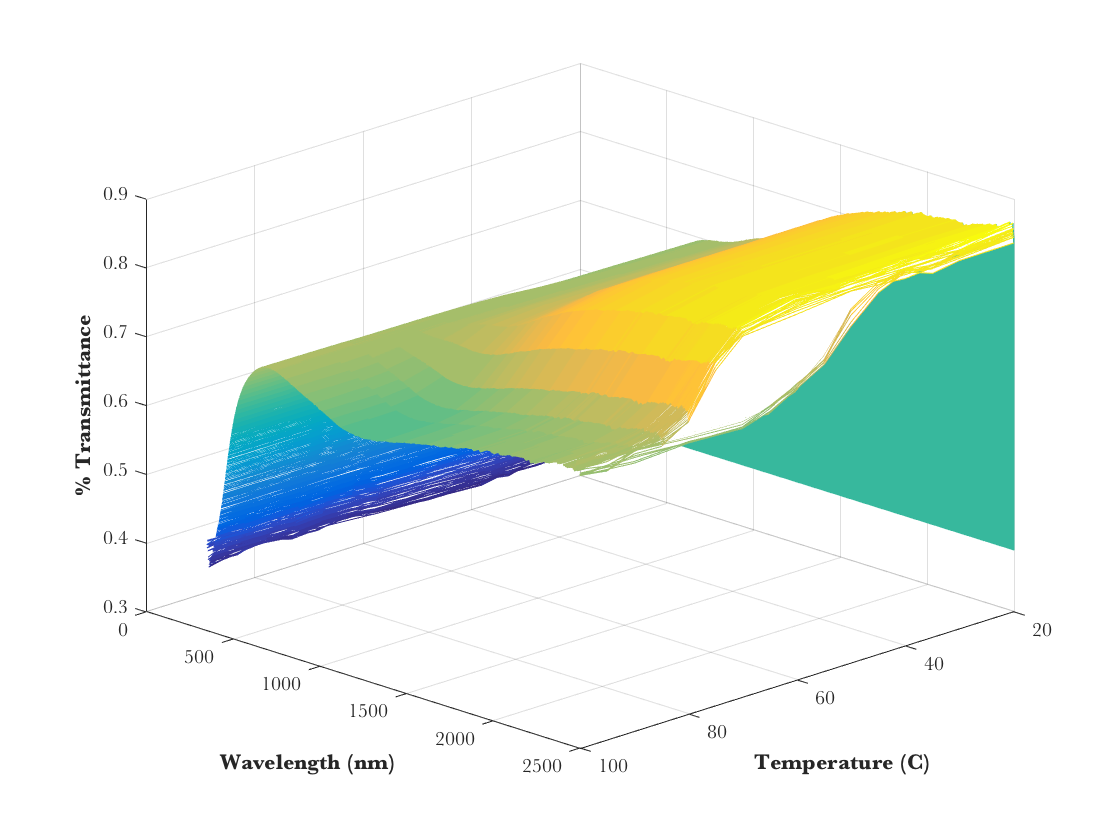}
    \caption{\textbf{Surface Plot}}
    \label{surf}
\end{subfigure}%
\begin{subfigure}{.5\textwidth}
  \centering
  \includegraphics[width=.95\linewidth]{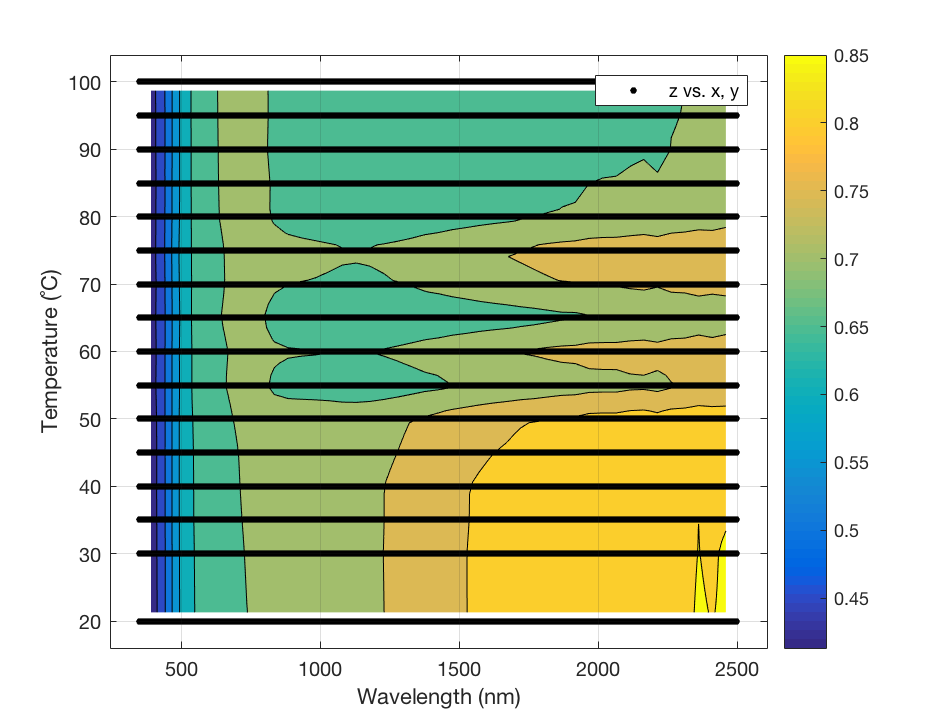}
  \caption{\textbf{Contour Map}}
  \label{contour}
\end{subfigure}
\caption{\textbf{0.93 VO$_x$-In$_2$O$_3$ 400$\degree$ C 66.7 ppm Spectra}}
\label{3d}
\end{figure*}

Of the samples that successfully formed VO$_2$, the sample annealed at $400\degree$C with $66.7$ ppm O$_2$ displayed both the highest luminous transmittance and the greatest transmittance modulation (See Figure \ref{3d}\subref{surf}, \ref{3d}\subref{contour} for a better visualization of this sample's thermochromic character). For all successful samples, a hysteresis curve was plotted for the transmittance at a wavelength of $2000$ nanometers to evaluate the transition in more detail. The heating and cooling sections were separated and the points were fit to a curve modeled by the following with T as percent transmittance and t as temperature.   

\begin{figure*}
\centering
\begin{subfigure}{.5\textwidth}
    \centering
    \includegraphics[scale=.5]{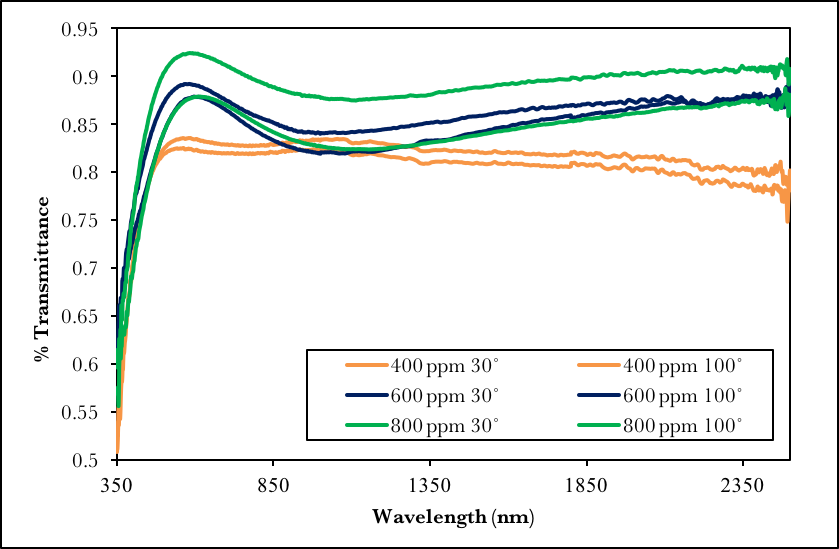}
    \caption{\textbf{0.93 VO$_x$-TiO$_2$ Spectra}}
    \label{Ti}
\end{subfigure}%
\begin{subfigure}{.5\textwidth}
  \centering
  \includegraphics[width=.95\linewidth]{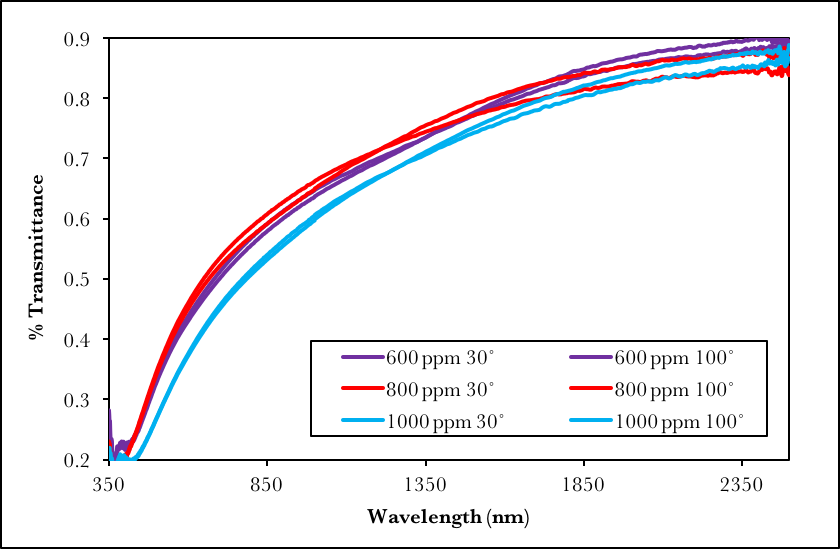}
  \caption{\textbf{0.93 VO$_x$-ZnO Spectra}}
  \label{ZnO}
\end{subfigure}
\caption{\textbf{VO$_x$-TiO$_2$ and VO$_x$-ZnO 375$\degree$ C Spectra}}
\label{spectra}
\end{figure*}


The accuracy of each curve fits was at least $0.9999$. The heating and cooling models for the $400\degree$C, $66.7$ ppm O$_2$ $0.93$ VO$_x$-In$_2$O$_3$ sample, the most optimal of the samples tested, had accuracies of $0.99999392$ and $0.99998832$, respectively. The model used provides valuable information about the hysteresis: transmittance modulation is indicated by $A$ and the difference in B values indicates the hysteresis width. The optimal sample (Figure \ref{hysteresis}) had a hysteresis width of $23.634\degree$C, revealing the amount by which the phase transition during cooling lags behind that during heating. 

\[T=\frac{A}{1+e^{\frac{t-B}{C}}}+D\]

The spectral transmittance data was objectively characterized by evaluating the transmittance spectra as the fraction of solar spectral irradiance transmitted through the film. A MATLAB script was used to process the data and to calculate the solar energy transmittance in both the on state (T$_{solar\_ON}$), when the film is heated above T$_{c}$, and the off state (T$_{solar\_OFF}$), when the film is below T$_{c}$. While T$_{solar}$ values were computed for wavelengths of light between 350 and 1500 nm, T$_{solar\_NIR\_ON}$, T$_{solar\_NIR\_OFF}$, T$_{lum\_ON}$, and T$_{lum\_OFF}$ values were determined for the NIR and luminous ranges of wavelengths when films were in the on and off states, respectively. 
    
 \[T_{sol}=\frac{\int E_{e,\lambda}(\lambda)T(\lambda) d\lambda}{\int E_{e,\lambda}(\lambda) d\lambda}\]


$\\$The equation for calculating solar energy transmittance is displayed above, with the limits of integration modified, corresponding to the range of wavelengths. This equation normalizes the transmittance percentages $[T(\lambda)]$ according to the solar spectral irradiance $[E_{e,\lambda}(\lambda)]$, a measure of the amount of heat transmitted at a given wavelength – to provide an objective quantification of ability of a certain film to filter heat.


\begin{figure*}
\centering
\begin{subfigure}{.5\textwidth}
  \centering
  \includegraphics[width=.95\linewidth]{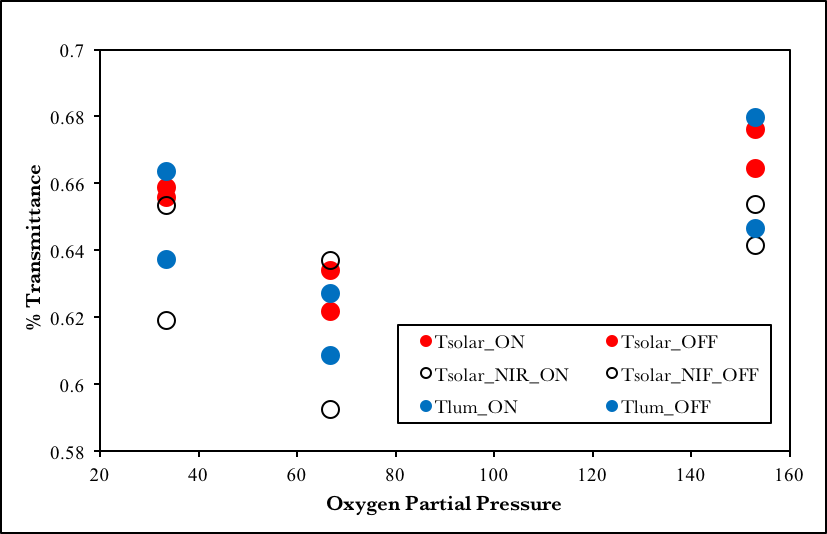}
  \caption{\textbf{375$\degree$C}}
  \label{SHGC375}
\end{subfigure}%
\begin{subfigure}{.5\textwidth}
  \centering
  \includegraphics[width=.95\linewidth]{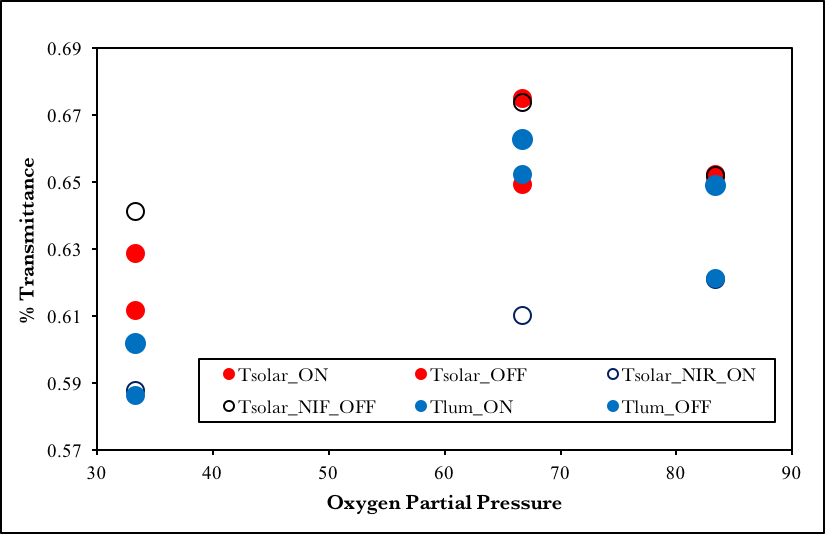}
  \caption{\textbf{400$\degree$C}}
  \label{SHGC400}
\end{subfigure}
\caption{\textbf{Solar Energy Transmittance in 0.93 VO$_x$-In$_2$O$_3$}}
\label{SHGC}
\end{figure*}

Certain general trends can be extrapolated from these data (Figure \ref{SHGC}). First, it is evident that the samples annealed at $400\degree$C yielded a greater solar energy transmittance both in the near infrared range of wavelengths and across the entire range of wavelengths. For both the set of samples annealed at $375\degree$C (Figure \ref{SHGC}\subref{SHGC375}) and that annealed at $400\degree$C (Figure \ref{SHGC}\subref{SHGC400}), there appears to be an optimal condition at which the transmittance modulation is maximized, since the central oxygen partial pressure corresponds to greater transmittance modulation than the first and last conditions. The luminous transmittance values are relatively high for all samples. This data combined with the transmittance data suggests that the sample of $0.93$ VO$_x$-In$_2$O$_3$ annealed at $400\degree$C with $66.7$ ppm O$_2$ is the most optimal, yielding a high solar energy modulation in the near infrared range while maintaining a relatively high luminous transmittance (Figure \ref{SHGC}\subref{SHGC400}). This demonstrates potential for real application—when doped with W$^{6+}$ ions, this synthesis method can produce a fully functional film.

The TiO$_2$ and ZnO samples were tested extensively as well, although neither sample successfully demonstrated modulation. The TiO$_2$ samples were annealed with $33.3$, $50$, $100$, $150$, $200$, $400$, $600$, $800$, and $1000$ ppm at $375\degree$C, while the ZnO samples were annealed at $83.3$, $250$, $400$, $600$, $800$, and $1000$ ppm at $375\degree$C.

These spectra suggest that as oxygen partial pressures during annealing increase, so does overall luminous transmittance since the luminous transmittance values for any wavelength are strictly increasing across the samples in the off state (Figure \ref{spectra}). These data also suggest that transmittance modulation may increase along with oxygen partial pressure, since the spectra appear to demonstrate greater switching. It is highly unexpected that $0.93$ VO$_x$-TiO$_2$ does not exhibit transmittance modulation under any conditions since TiO$_2$ has such a large dielectric constant (Figure \ref{spectra}\subref{Ti}). Instead, a likely explanation is that VO$_x$-TiO$_2$ requires a greater oxygen partial pressure, perhaps due to the oxygen diffusivity of TiO$_2$. This is a key area for future work, since TiO$_2$ demonstrates significant potential as a diluent in a VO$_2$ nanocomposite film, although it may require oxygen partial pressure beyond 1000 ppm for perceivable transmittance modulation to occur; this could not be tested due to limitations in the equipment available, as the O$_2$ without dilution did not exceed $1000$ ppm. 

The VO$_x$-ZnO spectra do not appear to display any sort of correlation; this is not surprising, since ZnO had a low dielectric constant, and it was therefore anticipated that it would not perform as well as TiO$_2$ (Figure \ref{spectra}\subref{ZnO}).

\begin{figure*}
\centering
\begin{subfigure}{.5\textwidth}
  \centering
  \includegraphics[width=.94\linewidth]{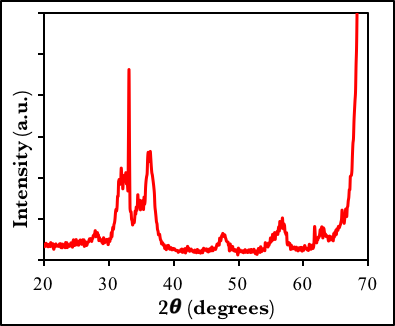}
  \caption{\textbf{0.93 600 ppm VO$_x$-ZnO XRD }}
  \label{ZnXRD}
\end{subfigure}%
\begin{subfigure}{.5\textwidth}
  \centering
  \includegraphics[width=.96\linewidth]{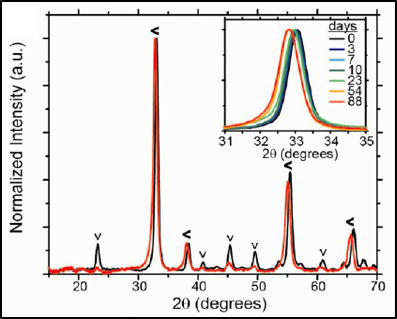}
  \caption{\textbf{VO$_2$ XRD }}
  \label{VO2XRD}
\end{subfigure}
\caption{\textbf{X-Ray Diffration Patterns}}
\label{XRD}
\end{figure*}
    

An XRD pattern was taken for a sample of VO$_x$-ZnO, known peaks were identified, and the remaining pattern was compared with that of VO$_2$. The peaks with intensities greater than $200$ a.u. can be attributed to the Silicon substrate. VO$_2$ peaks occur at approximately $2\theta = 27, 42, 45, 49, 53, 59, 69$ and a slightly more intense peak occurs at $2\theta = 37$. While some of these match the pattern, likely due to the ZnO, many expected peaks cannot be seen, such as the stronger peak at $2\theta = 37$ and the weaker peaks at $2\theta = 49$ and $53$. This indicates that either VO$_2$ is not present and another oxide of vanadium was created, or that VO$_2$ is present in such trace quantities that it is imperceptible on the XRD, and would consequently have an almost imperceptible transmittance modulation. Given the dynamic nature of Vanadium oxides and the precise conditions required to successfully synthesize VO$_2$, the former explanation appears to be more likely.

Although the VO$_x$-TiO$_2$ and the VO$_x$-ZnO samples failed to demonstrate transmittance modulation, the data collected provides valuable information which will be crucial in future experiments.  The VO$_x$-TiO$_2$ exhibited luminous transmittance at unprecedented levels, far greater than those of the VO$_x$-In$_2$O$_3$ samples, and while substantial transmittance modulation was not observed with an O$_2$ partial pressure below $1000$ ppm, it is expected that modulation will be observed for higher partial pressures (Figure \ref{In}, \ref{spectra}\subref{Ti}). This is largely due to the interference pattern and observed shape of the spectra, which begin to take a very similar form to those of successful samples, with increasing transmittance modulation as partial pressures increase. Much like the VO$_x$-In$_2$O$_3$ spectra in monoclinic phase (at $30\degree$C), the 800 ppm VO$_x$-TiO$_2$ spectra reveals a peak in the visible light region (between $500$ and $700$ nm) and a dip as the wavelengths approach the near-infrared region. This prompts further questions about the unique nature of the VO$_x$-TiO$_2$ interaction, due to the interesting results of this experiment, the computationally predicted success of TiO$_2$ as a host matrix (as per \cite{li}), and the generally low O$_2$ partial pressures observed in the literature for VO$_2$ films. The results of this experiment are valuable in the direct implications of the successful transmittance modulation of the VO$_x$-In$_2$O$_3$ spectra, as well as the indirect implications of the unexpected, unique data collected for the VO$_x$-TiO$_2$ spectra \footnote{Further results can be found in Appendix A: Sample Collection}.

\section{Conclusion}
This investigation confirms the general hypothesis that diluting VO$_2$ nanocrystalline films with white or clear metal oxide nanocrystals will increase luminous transmittance while preserving thermochromism. The VO$_x$-In$_2$O$_3$ spectra provide clear evidence of this, although more data would strengthen this conclusion. Furthermore, the VO$_x$-TiO$_2$ films partly substantiate the hypothesis, as it is apparent that there is a positive correlation between oxygen partial pressure and overall luminous transmittance but not transmittance modulation, since the modulation demonstrated by the VO$_x$-TiO$_2$ films in the experiment is not significant enough to be valuable in any application. 

The evidence presented is somewhat consistent with the computed correlation between highly dielectric materials and high thermochromism, since it is possible that were higher oxygen partial pressure to be tested, the VO$_x$-TiO$_2$ sample would continue the observed correlation of increasing transmittance modulation. In general, the In$_2$O$_3$ functioned the best as a diluting agent, since the other two samples ultimately failed to demonstrate meaningful transmittance modulation. Regarding the trends in annealing conditions, it is clear that samples performed better at $400\degree$C than at $375\degree$C. Additionally, the samples annealed at $66.7$ ppm O$_2$ appeared to performed better than those annealed at $33.3$ ppm O$_2$, although the samples appeared at $50$ ppm inexplicably failed; this sample’s failure represents an interesting area to explore further.

Based on a comprehensive cost analysis, the VO$_x$-In$_2$O$_3$ films introduced cost only \$1.71 / ft$^2$, which is almost 4x cheaper than average window films and 7x cheaper than specialty films, saving the average building approximately \$64,000 on costs. Solution-processing further decreases costs while enabling a scalable and versatile synthesis protocol. Widespread use of these films would save an estimated 5\% of the annual US energy budget, amounting to approximately 2 billion dollars.

The experiment was not wholly free from error as sources of bias still existed. Given the limited time available for this investigation, assessments of repeatability represented a significant limitation. Unique conditions were only tested in batches of two; although most films produced transmittance spectra almost identical to their counterparts, additional tests would be beneficial to further validate the legitimacy of the conclusions drawn from the data presented. Another potential source of error could be the unintentional oxidation of the films. Although they were stored in the oxygen-free vacuum dessicator when not in use, the films were often kept out during experimentation which may have influenced results, albeit not to a significant extent. The testing in the Linkam Cell, which is not an oxygen-free environment as it is not vacuum sealed, exposed the films to oxygen for up to hours at a time; however, this was largely unavoidable. A third, and more potent, error source might be the inconsistency of film thickness across the substrate. Since the spectrophotometer only collects transmittance data for a certain point on the film, which is calibrated manually, having films of uneven thickness would likely make the data inconsistent, since measuring a thicker region of film would yield lower transmittance percentages. Although there is no reason to assume that film thickness would be drastically inconsistent, minor variations in uniformity are inevitable due to the insufficient precision of the spin coater. These sources of error were minimized during the experimental process, but a limited amount of error was inevitable, as is in any complex experiment.

The equipment undeniably contributed error, although the magnitude of this error cannot be objectively determined due to the nature of the experimental data: graphs across a range of wavelengths were collected rather than discrete measurements at a few points. Light contamination represents one likely source of error for the spectrophotometer, as measurements taken in a lighted room, thus ambient light may have entered the Linkam Cell. However, this is a relatively minor concern, since this would have caused a greater transmittance to be perceived across all wavelengths, while preserving relative transmittance between different films, the measurement in which this study is primarily interested.

In future, more experimentation regarding the VO$_x$-TiO$_2$ and VO$_x$-ZnO samples would be valuable to further substantiate the observations made and to minimize error involved in the data collection.  Additionally, testing more metal oxide diluents, particularly higher dielectric metal oxides, would greatly benefit this research, as would the testing of a wider range of partial pressures and temperatures. In continuation of this project, VO$_x$-In$_2$O$_3$ and VO$_x$-TiO$_2$, will be tested with a wide variety of variables and each sample will be tested multiple times to ensure accuracy. It would additionally be of value to computationally model the film chemistry to further validate the conclusions drawn from the results of the experimentation. Another concern is the Tungsten doping which has been confirmed in the literature to decrease the transition temperature of VO$_2$; this effect was demonstrated on epitaxial thin films and could potentially be altered by the introduction of a host nanocrystalline matrix, thus the incorporation of doped VO$_2$ represents an avenue for future research.

This investigation poses numerous questions relating to the nature of the diluent and its influence on the optical properties of the nanocomposite film. Most notably, it interrogates the physical phenomena responsible for the lack of significant transmittance modulation exhibited by VO$_x$-TiO$_2$ even at high oxygen partial pressures. Furthermore, what is the minimum oxygen partial pressure required to elicit transmittance modulation from VO$_x$-TiO$_2$? Why are there no strict trends in thermochromism of VO$_x$-In$_2$O$_3$ across O$_2$ partial pressure? 
    
In conclusion, this work demonstrates the possibility for synthesis of low-cost, solution-processed, thermochromic VO$_2$ nanocomposite films with both high transmittance modulation and luminous transmittance with VO$_x$-In$_2$O$_3$ as the diluent. Our investigation demonstrates a significant increase in luminous transmittance of VO$_2$ nanocomposite films without any drastic impact on thermochromic character; this has critical implications, and brings this technology much closer to its societal applications. This study represents a crucial development in film chemistry and paves the way for further application of VO$_2$ nanocomposite films in chromogenic fenestration.


\bibliographystyle{alpha}

\end{document}